\documentclass[3p,10pt,authoryear]{elsarticle}

\usepackage{amsmath,bm,epsfig}
\newcommand{\B}[1]{{\bm{#1}}}
\newcommand{\C}[1]{{\mathcal{#1}}}
\newcommand{\pa}{\partial}
\usepackage[latin1]{inputenc}
\usepackage{natbib}

\begin{document}

\begin{frontmatter}

\title{The $1/r$ singularity in weakly nonlinear fracture mechanics}

\author{Eran Bouchbinder, Ariel Livne and Jay Fineberg}
\address{Racah Institute of Physics, Hebrew University of Jerusalem, Jerusalem 91904, Israel}

\begin{abstract}
Material failure by crack propagation essentially involves a concentration of large displacement-gradients near a crack's tip, even at scales where no irreversible deformation and energy dissipation occurs. This physical situation provides the motivation for a systematic gradient expansion of general nonlinear elastic constitutive laws that goes beyond the first order displacement-gradient expansion that is the basis for linear elastic fracture mechanics (LEFM). A weakly nonlinear fracture mechanics theory was recently developed by considering displacement-gradients up to second order. The theory predicts that, at scales within a dynamic lengthscale $\ell$ from a crack's tip, significant $\log{r}$ displacements and $1/r$ displacement-gradient contributions arise. Whereas in LEFM the $1/r$ singularity generates an unbalanced force and must be discarded, we show that this singularity not only exists but is {\em necessary} in the weakly nonlinear theory. The theory generates no spurious forces and is consistent with the notion of the autonomy of the near-tip nonlinear region. The J-integral in the weakly nonlinear theory is also shown to be path-independent, taking the same value as the linear elastic J-integral. Thus, the weakly nonlinear theory retains the key tenets of fracture mechanics, while providing excellent quantitative agreement with measurements near the tip of single propagating cracks. As $\ell$ is consistent with lengthscales that appear in crack tip instabilities, we suggest that this theory may serve as a promising starting point for resolving open questions in fracture dynamics.
\end{abstract}


\end{frontmatter}

\section{Introduction}
\label{intro}

The dynamics of propagating cracks, and their accompanying instabilities \citep{99FM,07LBDF}, remain a fundamental and challenging scientific problem.
The most developed theoretical approach to fracture dynamics, Linear Elastic Fracture Mechanics (LEFM) \citep{98Fre, 99Bro}, is based on the assumption that deviations from a linear elastic constitutive material behavior takes place only in a small region near the tip of a crack. The major prediction of this theoretical framework is that outside of this small nonlinear near-tip region, but not very far from it, the displacement-gradient (and stress) fields are characterized by a $1/\sqrt{r}$ singularity, where $r$ is measured from the crack's tip. The intensity of the singularity is quantified by the stress intensity factor, $K$, that appears as a pre-factor in the singular fields and depends on the applied loadings and geometric configuration in a given problem. A corollary of this basic result is that the crack tip opening displacement (CTOD) is predicted to be parabolic, where the curvature of the parabola is a function of $K$.

A central concept in LEFM is that of the {\em autonomy} of the near-tip nonlinear zone \citep{68Rice_a, 98Fre, 99Bro}. The basic idea is that the mechanical state within the near-tip nonlinear zone, which is surrounded by the $1/\sqrt{r}$ singular fields (the ``K-fields''), is determined uniquely by the value of $K$, but is otherwise independent of the applied loadings and the geometric configuration in a given problem. This implies, for example, that systems with the same $K$, but with different applied loadings and geometric configurations, will be in the same mechanical state within the near-tip nonlinear zone.

One of the most useful applications of the $1/\sqrt{r}$ singular fields is the calculation of the energy flow into the nonlinear near-tip zone during crack propagation. It is calculated using a path-independent integral, the J-integral \citep{68Rice_b, 98Fre}, that is evaluated within the range of dominance of the $1/\sqrt{r}$ singular fields. This energy flux, which is quadratic in $K$, is eventually dissipated at smaller scales near the tip of the crack.

The $1/\sqrt{r}$ singular fields, however, imply that at some distance from the crack's tip the small displacement-gradient assumption of LEFM must break down when higher order displacement-gradients become non-negligible. This breakdown of LEFM is expected to take place prior to the intervention of irreversible deformation and energy dissipation, and thus should be studied in the framework of nonlinear elasticity. This physical situation motivates a controlled expansion of a general nonlinear elastic material constitutive law in powers of displacement-gradients, beyond the common small displacement-gradient approximation of LEFM. Such an approach was taken recently, when a weakly nonlinear fracture mechanics theory was developed by deriving the leading nonlinear elastic correction, up to second order displacement-gradients, to the $1/\sqrt{r}$ singular fields of LEFM \citep{08LBF,08BLF}. It is important to note that this approach is entirely general as any material exhibits nonlinear elastic response to some degree. Moreover, it entails no new constitutive assumptions and/or the introduction of a new intrinsic material lengthscale.

The new theory was tested against direct measurements of the deformation near the tip of propagating cracks \citep{08LBF} and was shown to be in excellent quantitative agreement with these experimental data \citep{08BLF}. The weakly nonlinear correction to LEFM includes displacement contributions proportional to $\log{r}$ and displacement-gradient contributions proportional to $1/r$. The $\log{r}$ contributions most clearly manifest themselves by a strong departure from a parabolic CTOD in the near vicinity of the crack tip. The $1/r$ strain contributions immediately imply that within the spatial range of validity of the weakly nonlinear solution, the displacement-gradient spatial variation is ``more divergent'' than the common $1/\sqrt{r}$ singularity.

The $1/r$ displacement-gradient singularity must be discarded in the framework of LEFM, since it generates an unbalanced force acting on a line encircling a crack's tip \citep{74Rice}. In this work we show that in the framework of the weakly nonlinear theory this singularity generates no such spurious force and is both necessary and physically acceptable. Furthermore, the condition that such a spurious force does not exist can be used to retain a key feature of LEFM, namely that of the {\em autonomy} of the near-tip nonlinear zone, implying that the weakly nonlinear solution depends only on the stress intensity factor $K$.
Finally, we calculate the weakly nonlinear J-integral, i.e. we expand the general nonlinear J-integral up to the leading nonlinear correction to LEFM, and show that it takes the same value as its LEFM counterpart. This result explicitly demonstrates the path-independence of the weakly nonlinear J-integral and, together with the notion of autonomy, shows that the weakly nonlinear theory is consistent with {\em all} of the common tenets of fracture mechanics.

The weakly nonlinear solution implies the existence of a lengthscale $\ell$ that is associated with the breakdown of LEFM near the tip of a propagating crack. $\ell$ is defined as the lengthscale at which the $1/r$ singular terms become non-negligible compared to the dominant $1/\sqrt{r}$ linear elastic singular terms and marks the onset of deformation-dependent material behavior \citep{96Gao, 03BAG, 06BG, 08BL}. We suggest that $\ell$ may be intimately related to crack tip instabilities that currently remain unexplained within the framework of LEFM. Therefore, the set of encouraging results obtained in this work leads us to suggest that the weakly nonlinear theory may be a promising starting point for unlocking a plethora of open questions in fracture dynamics. This line of investigation is close in spirit to other recently proposed approaches that highlight the role of nonlinear elasticity near the tip of propagating cracks \citep{03BAG,06BG}.

The structure of this paper is as follows. In Sect. \ref{details} we describe the recently derived weakly nonlinear solution \citep{08BLF} and include a detailed derivation of the theory. In Sect. \ref{free} we discuss the difference between LEFM and the weakly nonlinear theory in relation to the $1/r$ singularity. We also demonstrate the autonomy of the near-tip nonlinear zone. This section culminates in a quasi-static weakly nonlinear solution. In Sect. \ref{comparison} we compare the resulting solution with direct measurements of the deformation near the tip of a crack propagating in a neo-Hookean material and demonstrate excellent agreement with experiments. In Sect. \ref{J} we show that the weakly nonlinear J-integral is path-independent, its value coinciding with the well-known LEFM result. Section \ref{summary} offers some concluding remarks.

\section{The details of the weakly nonlinear solution}
\label{details}

As noted in the introduction, nonlinear material response at the large strains near a crack's tip in general and the experimental results of \citet{08LBF} in particular,
motivate us to formulate a nonlinear elastic dynamic fracture
problem under plane stress conditions. In this section we derive, in detail, the weakly nonlinear theory, which was previously outlined in the results of \citet{08BLF}. Consider the deformation
field $\B \phi$, which is assumed to be a continuous, differentiable
and invertible mapping between a reference configuration $\B x$ and
a deformed configuration $\B x'$ such that
\begin{equation}
\label{deform}
\B x' = \B \phi(\B x)=\B x+\B u(\B x) \ ,
\end{equation}
where $\B u(\B x)$ is the displacement field.
The deformation gradient tensor $\B F$ is
defined as
\begin{equation}
\label{def_grad}
\B F = \nabla \B \phi \ ,
\end{equation}
or explicitly as $F_{ij}\!=\!\delta_{ij}+\pa_j u_i$, where $i,j$ denote Cartesian components. The first Piola-Kirchhoff stress
tensor $\B s$, that is work-conjugate to the deformation gradient
$\B F$, is given as
\begin{equation}
\label{s}
\B s = \frac{\pa U(\B F)}{\pa {\B F}} \ ,
\end{equation}
where $U(\B F)$ is the strain energy in the deformed configuration per unit volume
in the reference configuration \citep{Holzapfel}. The momentum balance equation is
\begin{equation}
\label{EOM}
\nabla \cdot \B s = \rho \pa_{tt}{\B \phi} \ ,
\end{equation}
where $\rho$ is the reference mass density. Under steady-state propagation
conditions we expect all of the fields to depend on $x$ and $t$
through the combination $x\!-\!vt$ and therefore
$\pa_t\!=\!-v\pa_x$. Here $x$ is the
propagation direction and $y$ is the
loading direction. The polar coordinate system $(r,\theta)$ that moves with the
crack tip is related to the rest frame by
$r\!=\!\sqrt{(x-vt)^2+y^2}$ and $\theta\!=\!\tan^{-1}[y/(x-vt)]$.
Thus, the traction-free boundary conditions on the crack's faces are
\begin{equation}
\label{BC}
s_{xy}(r,\theta\!=\!\pm\pi)\!=\!s_{yy}(r,\theta\!=\!\pm\pi)=0 \ .
\end{equation}

We now perform a controlled expansion of the form
\begin{equation}
\label{expansion}
\B u(r,\theta) \simeq \epsilon \B u^{(1)}(r,\theta)+\epsilon^2 \B u^{(2)}(r,\theta) + \C O(\epsilon^3) \ ,
\end{equation}
where $\epsilon$ is a measure of the magnitude of displacement-gradients. The expansion is perturbative in the sense that $\C O(\epsilon^3)$
contributions to the deformation fields are neglected and asymptotic in the sense that we consider a region near the crack tip where $\C O(\epsilon^2)$ contributions are non-negligible compared to the leading $\C O(\epsilon)$ contributions in this region. Equation (\ref{expansion}) implies that we consider first (linear) and second (quadratic) orders of elasticity.
The order $\epsilon$ problem is the standard LEFM one
\begin{equation}
\mu\nabla^2{\B u^{(1)}}+(\tilde \lambda + \mu) \nabla(\nabla\cdot{\B u^{(1)}})=\rho\ddot{\B u}^{(1)} \ ,
\label{Lame}
\end{equation}
with the traction-free boundary conditions on the crack's faces at $\theta\!=\!\pm\pi$
\begin{eqnarray}
&&\label{BC1} \frac{\pa_\theta u_x^{(1)}}{r}+\pa_ru_y^{(1)}=0,\nonumber\\
&&(\tilde\lambda+ 2\mu) \frac{\pa_\theta u_y^{(1)}}{r}+\tilde \lambda \pa_r u_x^{(1)}=0 \ .
\end{eqnarray}
Here $\tilde \lambda$ is the plane-stress first Lam\'e coefficient and $\mu$ is the shear modulus.
The near-tip asymptotic solution under Mode I steady-state propagation conditions is known to be \citep{98Fre}
\begin{eqnarray}
\epsilon u_x^{(1)}(r, \theta;v)&=&\frac{K_I \sqrt{r}}{4\mu\sqrt{2\pi}}\Omega_x(\theta;v)+\frac{(\tilde\lambda+2\mu)Tr\cos\theta}{4\mu(\tilde\lambda+\mu)},\nonumber\\
\label{firstO}
\epsilon u_y^{(1)}(r,\theta;v)&=&\frac{K_I\sqrt{r}}{4\mu\sqrt{2\pi}}\Omega_y(\theta;v)-\frac{\tilde\lambda Tr\sin\theta}{4\mu(\tilde\lambda+\mu)}.
\end{eqnarray}
Here $K_I$ is the Mode I stress intensity factor (denoted in the introduction by $K$) and $T$ is a
constant known as the ``T-stress'' \citep{98Fre}. Note that these parameters
cannot be determined by the asymptotic analysis as they depend on
the {\em global} crack problem. $\B \Omega(\theta;v)$ is a known
universal function \citep{98Fre, 08BLF}. $\epsilon$ in
Eq. (\ref{expansion}) can be now defined explicitly as
\begin{equation}
\label{epsilon}
\epsilon \equiv \frac{K_I}{4\mu \sqrt{2\pi \ell(v)}} \ ,
\end{equation}
where $\ell(v)$ is a velocity-dependent length-scale. $\ell(v)$ defines the scale
where order $\epsilon^2$ contributions become non-negligible compared to order $\epsilon$ contributions, which is a property of the nonlinear elastic constitutive behavior of a given material. This is a ``dynamic'' (as opposed to ``intrinsic'') length-scale that marks the onset of deviations from a linear elastic constitutive behavior.

The order $\epsilon^2$ Mode I problem has the following form
\begin{equation}
\mu\nabla^2{\B u^{(2)}}+(\tilde\lambda+\mu)\nabla(\nabla\cdot{\B u^{(2)}})+
\frac{\mu\ell\B g(\theta;v)}{r^2}=\rho\ddot{\B u}^{(2)}\ .
\label{secondO}
\end{equation}
To this order the traction-free boundary conditions at $\theta\!=\!\pm\pi$ now become
\begin{eqnarray}
&&\label{BC2} \frac{\pa_\theta u_x^{(2)}}{r}+\pa_ru_y^{(2)}=0,\nonumber\\
&&(\tilde\lambda+ 2\mu) \frac{\pa_\theta u_y^{(2)}}{r}+\tilde \lambda \pa_r u_x^{(2)}+\frac{\mu\ell\kappa(v)}{r}=0.
\end{eqnarray}
Here $\B g(\theta;v)$ and $\kappa(v)$ are universal functions that depend on the first and second order elastic moduli.
Generic quadratic nonlinearities of
the form $\pa(\pa u^{(1)} \pa u^{(1)})$ result in
an effective body-force $\propto\!r^{-2}$ in Eq. (\ref{secondO}).
Note that corrections proportional to $K_I T r^{-3/2}/\mu$ were neglected in Eq. (\ref{secondO}). These corrections lead to displacement contributions that vary as $r^{1/2}$ and thus result in a few percent variation of the {\em apparent} $K_I$ with the angle $\theta$. This may indeed be observed, cf. Figs. \ref{fit} and \ref{fit1}; see also \citet{08LBF} and \citet{08BLF}.

The solution of the order $\epsilon^2$ Mode I problem posed by Eqs. (\ref{secondO})-(\ref{BC2}) is obtained in two steps. First we obtain a particular solution $\ell \B \Upsilon(\theta;v)$ of Eq. (\ref{secondO}). This solution is r-independent and results from solving a vectorial linear ordinary differential equation in $\theta$ that depends on $\B g(\theta;v)$ and thus on the second order elastic moduli. For the particular case of a neo-Hookean material, $\B \Upsilon(\theta;v)$ (for a general $v$) was obtained in \citet{08BLF}. Since $\ell \B \Upsilon(\theta;v)$ {\em does not} satisfy the boundary conditions of Eq. (\ref{BC2}), we have to add to it a solution of the homogeneous equation, obtained by omitting the inhomogeneous term in Eq. (\ref{secondO}). The resulting equation is just the Lam\'e equation of (\ref{Lame}). The homogeneous solution is selected such that the sum of the particular and homogeneous solutions satisfies the boundary conditions of Eq. (\ref{BC2}). Denoting the homogeneous solution by $\ell \tilde{\B u}$, we obtain
\begin{eqnarray}
\label{solution}
\epsilon^2u_x^{(2)}(r,\theta;v)&=&\left(\frac{K_I}{4\mu \sqrt{2\pi}}\right)^2\left[\tilde u_x(r,\theta;v)+\Upsilon_x(\theta;v)\right],\nonumber\\
\epsilon^2u_y^{(2)}(r,\theta;v)&=&\left(\frac{K_I}{4\mu
\sqrt{2\pi}}\right)^2\left[\tilde u_y(r,\theta;v)+\Upsilon_y(\theta;v)\right].
\end{eqnarray}

To obtain $\tilde{\B u}$ we follow a standard procedure \citep{98Fre, 99Bro}. Using the Helmholtz decomposition
\begin{equation}
\label{Helmholtz}
\tilde{\B u}=\nabla\varphi + \nabla \times \B \psi,
\end{equation}
we obtain
\begin{equation}
\label{wave1}
\alpha^2_d\partial_{x}^2\varphi+\partial_{y}^2\varphi=0,\quad
\alpha^2_s\partial_{x}^2\psi+\partial_{y}^2\psi=0 \ ,
\end{equation}
where $\B \psi = \psi \B z$ ($\B z$ is a unit vector perpendicular to the xy-plane).
Here $\alpha^2_{d,s}\equiv1-v^2/c_{d,s}^2$, where $c_{d,s}$ are the dilatational and shear wave speeds respectively and $\varphi,\psi$ are the commonly used displacement potentials \citep{98Fre}.
These equations can be rewritten as two Laplace's equations in the complex variables
\begin{equation}
\zeta_d=x+i\alpha_d y \equiv r_{d}e^{i\theta_{d}}\quad\hbox{and}\quad \zeta_s=x+i \alpha_s y \equiv r_{s}e^{i\theta_{s}},
\end{equation}
which are coupled only on the boundaries. Note that $\tan{\theta_{d,s}}\!=\!\alpha_{d,s}\tan{\theta}$ and $r_{d,s}\!=\!r\sqrt{1-(v\sin\theta/c_{d,s})^2}$.
The most general solution of Eqs. (\ref{wave1}) is given as
\begin{equation}
\label{analytic}
\varphi(r, \theta)=\Re\{R(\zeta_d) \},\quad \psi(r, \theta)=\Im\{Q(\zeta_s)\} \ ,
\end{equation}
where $R$ and $Q$ are two analytic functions. The crucial point for our discussion is that the second boundary condition of Eq. (\ref{BC2}) contains a term that varies as $r^{-1}$. Therefore, the solution must be one that has the property that $\nabla \tilde{\B u} \propto r^{-1}$. The solution with this property is given by
\begin{equation}
\label{analytic}
R(\zeta_d)=A\zeta_d\log{\zeta_d},\quad Q(\zeta_s)=B\zeta_s\log{\zeta_s} \ .
\end{equation}
Using the relations (cf. Eq. (\ref{Helmholtz}))
\begin{equation}
\tilde u_x=\partial_x\varphi+\partial_y\psi,\quad \tilde u_y=\partial_y\varphi-\partial_x\psi \ ,
\end{equation}
we obtain
\begin{eqnarray}
\label{solutionA}
\tilde u_x(r,\theta,v)&=&A\log{r}+\frac{A}{2}\log{\left[1-\frac{v^2\sin^2\theta}{c_d^2} \right]}+B\alpha_s\log{r}+\frac{B \alpha_s}{2}\log{\left[1-\frac{v^2\sin^2\theta}{c_s^2} \right]},\nonumber\\
\tilde u_y(r,\theta,v)&=&-A\alpha_d\theta_d-B\theta_s \ .
\end{eqnarray}
When this solution is substituted in the boundary conditions of Eq. (\ref{BC2}), a {\em single} relation between $A$ and $B$ is obtained \citep{08BLF}. The remaining free parameter, say $B$, was previously determined in \citet{08BLF} by fitting the solution to the experimental data. In the next section we will determine this free parameter theoretically. This result shows that no free parameters exist in the asymptotic solution in addition to $K_I$ (and possibly $T$). This is in accord with the concept of the autonomy of the nonlinear near-tip region.

\section{The autonomy of the near-tip nonlinear zone}
\label{free}

In this section we directly address the difference between LEFM and the weakly nonlinear theory in relation to the existence of the $1/r$ singularity. We will show how this difference is used to retain the autonomy of the near-tip nonlinear zone.
Let us now, for concreteness, consider the neo-Hookean plane-stress elastic strain energy functional \citep{83KS} that describes the constitutive behavior of the elastomer gels used in the experiments of \citet{08LBF}
\begin{equation}
\label{NH} U(\B F)= \frac{\mu}{2}\left[F_{ij}F_{ij}+\lambda^{2}-3\right] \ ,
\end{equation}
where the out of plane stretch $\lambda$ (not to be confused with the plane-stress first Lam\'e coefficient $\tilde \lambda$) is given by
\begin{equation}
\label{stretch}
\lambda = \det(\B F)^{-1} \ .
\end{equation}
The last relation corresponds to the incompressibility condition, which implies a single elastic modulus $\mu$. Incompressibility entirely determines the second linear elastic modulus to be $\tilde \lambda = 2 \mu$, which is equivalent to the incompressible linear elastic Poisson ration $\nu=1/2$.
For later use we write $\lambda(r,\theta)$ of Eq. (\ref{stretch}) explicitly as
\begin{equation}
\label{lamb}
\lambda(r,\theta) = \frac{r}{\pa_r \phi_x \pa_\theta \phi_y-\pa_r \phi_y
\pa_\theta \phi_x} \ .
\end{equation}
The equations of motion corresponding to this energy functional, derived using Eqs. (\ref{s})-(\ref{EOM}), become
\begin{eqnarray}
\label{EOM1}
\mu\nabla^2\phi_x\!\!&+&\!\!\frac{\mu}{r}\left[\pa_\theta\lambda^3\pa_r \phi_y\!-\!\pa_r\lambda^3\pa_\theta \phi_y \right]\!=\!\rho v^2 \pa_{xx}\phi_x,\nonumber\\
\mu\nabla^2\phi_y\!\!&+&\!\!\frac{\mu}{r}\left[\pa_r\lambda^3\pa_\theta
\phi_x\!-\!\pa_\theta\lambda^3\pa_r \phi_x \right]\!=\!\rho v^2
\pa_{xx}\phi_y,
\end{eqnarray}
where the assumption of steady-state propagation condition was used.
The traction-free boundary conditions on the crack's faces, cf. Eq. (\ref{BC}), are
\begin{eqnarray}
\label{BCpolar}
\left[\frac{1}{r}\pa_\theta\phi_x+\lambda^3\pa_r\phi_y\right]_{\theta=\pm\pi}&=&0, \nonumber\\
\left[\frac{1}{r}\pa_\theta\phi_y-\lambda^3\pa_r\phi_x\right]_{\theta=\pm\pi}&=&0 \ .
\end{eqnarray}

In order to derive the functions $\B g(\theta;v)$ and $\kappa(v)$ in Eqs. (\ref{secondO})-(\ref{BC2}) we expand $\lambda^3(r,\theta)$, where $\lambda$ is given in Eq. (\ref{lamb}), to second order in $\epsilon$ (see Appendix).
Substituting this expansion (see Eq. (\ref{lambdaExpansion}) in the Appendix) into Eqs. (\ref{EOM1})-(\ref{BCpolar}) and using the first order solution of Eq. (\ref{Lame}) we obtain explicit expressions for the functions  $\B g(\theta;v)$ and $\kappa(v)$ in Eqs. (\ref{secondO})-(\ref{BC2}).

To elucidate the mathematical procedure to follow and in order to derive analytical results, we now restrict ourselves to the quasi-static limit $v \to 0$.
In this limit we obtain
\begin{eqnarray}
\label{bodyforceAngular}
g_x(\theta;v\!=\!0) \!&=&\!\frac{1}{12}\left[-85\cos\theta+208\cos(2\theta)-27\cos(3\theta)\right],\nonumber\\
g_y(\theta;v\!=\!0) \!&=&\! \frac{1}{12}\left[-79\sin\theta+208\sin(2\theta)-27\sin(3\theta)\right],\nonumber\\
\kappa(v=0) \!&=&\! -\frac{64}{3} \ .
\end{eqnarray}
Using $\B g(\theta,0)$ of Eq. (\ref{bodyforceAngular}), we obtain the components of the particular (inhomogeneous) part of the solution appearing in Eq. (\ref{solution})
\begin{eqnarray}
\label{particular}
\Upsilon_x(\theta;0)\!&=&\! - \frac{103}{48} \cos{\theta} +\frac{26}{15} \cos{(2\theta)} - \frac{3}{16} \cos{(3\theta)},\nonumber\\
\Upsilon_y(\theta;0)\!&=&\! - \frac{61}{48}\sin{\theta} +\frac{26}{15} \sin{(2\theta)} - \frac{3}{16} \sin{(3\theta)}.
\end{eqnarray}

The quasi-static limit of the homogeneous part of solution appearing in Eq. (\ref{solution}) is a bit subtle. Both the parameters $A$ and $B$ in Eq. (\ref{analytic}) {\em diverge} in the limit $v \to 0$ according to
\begin{eqnarray}
\label{diverge}
A&\simeq&\alpha + \frac{\gamma c_s^2}{v^2},\nonumber\\
B&\simeq&\beta - \frac{\gamma c_s^2}{v^2} \ .
\end{eqnarray}
A similar divergence is encountered in the asymptotic LEFM result through the appearance of the Rayleigh function \citep{98Fre, 99Bro}.
Therefore, simply substituting $v\!=\!0$ in Eq. (\ref{solution}) gives a {\em wrong} result in which there is only a single parameter ($A+B$) in the solution.
To circumvent this problem, we expand the homogeneous part of the solution appearing in Eq. (\ref{solutionA}) to second order in $v/c_s$
\begin{eqnarray}
&&\tilde u_x(r,\theta;0)\simeq(A+B)\log{r}+\left(-\frac{1}{2}B\log{r}-\frac{1}{8}A \sin^2\theta-\frac{1}{2}B\sin^2\theta \right)\frac{v^2}{c_s^2}+\C O(v^3),\nonumber\\
&&\tilde u_y(r,\theta;0)\simeq-(A+B)\theta+\left(\frac{1}{8}A\theta+\frac{1}{16}A \sin(2\theta)+\frac{1}{4}B\sin(2\theta) \right)\frac{v^2}{c_s^2}+\C O(v^3),
\end{eqnarray}
where we used $c_d\!=\!2c_s$ in Eq. (\ref{solutionA}), the appropriate relation for an incompressible material under plane stress conditions.
Using Eq. (\ref{diverge}), the limit $v \to 0$ now yields
\begin{eqnarray}
\label{QS2nd}
\tilde u_x(r,\theta;0)&=&a\log{r}+b\left(\log{r}+\frac{3}{4}\sin^2\theta \right),\nonumber\\
\tilde u_y(r,\theta;0)&=&-a\theta+b\left(\frac{\theta}{4}-\frac{3}{8}\sin(2\theta) \right)\ ,
\end{eqnarray}
where
\begin{equation}
a\equiv \alpha + \beta\quad\hbox{and}\quad b\equiv \frac{\gamma}{2}.
\end{equation}

We now focus on determining the parameters $a$ and $b$ in Eq. (\ref{QS2nd}). One constraint on these parameters is obtained from the boundary conditions of Eq. (\ref{BC2}), with $\kappa(0)$ of Eq. (\ref{bodyforceAngular}). This implies
\begin{eqnarray}
\label{a}
a = -\frac{1}{15}.
\end{eqnarray}
The determination of the second parameter, $b$, follows from an interesting feature of the $1/r$ singularity of the displacement-gradients. As in quasi-static LEFM, the $1/r$ singular term in a Williams expansion \citep{57Will} generates a net force transmitted across any line surrounding the crack tip that acts in the propagation direction, when $b$ is not properly selected \citep{74Rice}. Since no such force is applied to the crack faces/tip and since inertia is negligible in the quasi-static limit, this apparent force is an unbalanced one that cannot physically exist. The condition that no net force is transmitted across a line surrounding the crack tip is uniquely satisfied in quasi-static LEFM by eliminating the $1/r$ strain singularity altogether \citep{74Rice}. This is {\em not} the case in the weakly nonlinear theory. To see this, we calculate the net force $\B f$ transmitted across a line surrounding the crack tip
\begin{eqnarray}
\label{spurious}
f_x &\propto&  \int_{-\pi}^{\pi} \left[s_{xx} n_x + s_{xy} n_y \right] r d\theta,\nonumber\\
f_y &\propto&  \int_{-\pi}^{\pi} \left[s_{yx} n_x + s_{yy} n_y \right]r d\theta \ ,
\end{eqnarray}
where the line is chosen to be a circle of radius $r$ whose outward normal is $\B n \!=\! (\cos\theta,\sin\theta)$. Note that the first Piola-Kirchhoff stress tensor, $\B s$, describes forces in the deformed configuration per unit area in the reference (un-deformed) configuration. Therefore, the integration in Eq. (\ref{spurious}) over circles in the reference configuration, gives the force $\B f$ on their images in the deformed configuration.

To evaluate the integrals in Eq. (\ref{spurious}), we derive the components of $\B s$ in Eq. (\ref{s}) for $U(\B F)$ of Eq. (\ref{NH})
\begin{equation}
\label{PK}
s_{ij}=\mu\left(\pa_j\phi_i-\lambda^3\epsilon_{ik}\epsilon_{jl}\pa_l\phi_k
\right) \ ,
\end{equation}
where $\epsilon_{ij}$ is the two-dimensional alternator. More explicitly, in term of the deformation field $\B \phi$ (cf. Eq. (\ref{deform})), we have
\begin{eqnarray}
\label{sij}
s_{xx}&=& \mu\left(\cos\theta\pa_r\phi_x-\lambda^3\sin\theta\pa_r\phi_y-\sin\theta\frac{\pa_\theta\phi_x}{r}-\lambda^3\cos\theta\frac{\pa_\theta\phi_y}{r}\right) ,\nonumber\\
s_{yx}&=& \mu\left(\cos\theta\pa_r\phi_y+\lambda^3\sin\theta\pa_r\phi_x-\sin\theta\frac{\pa_\theta\phi_y}{r}+\lambda^3\cos\theta\frac{\pa_\theta\phi_x}{r}\right),\nonumber\\
s_{xy}&=& \mu\left(\sin\theta\pa_r\phi_x+\lambda^3\cos\theta\pa_r\phi_y+\cos\theta\frac{\pa_\theta\phi_x}{r}-\lambda^3\sin\theta\frac{\pa_\theta\phi_y}{r}\right),\nonumber\\
s_{yy}&=& \mu\left(\sin\theta\pa_r\phi_y-\lambda^3\cos\theta\pa_r\phi_x+\cos\theta\frac{\pa_\theta\phi_y}{r}+\lambda^3\sin\theta\frac{\pa_\theta\phi_x}{r}\right).
\end{eqnarray}

To expand $\B s$ in powers of $\epsilon$, we need to expand $\B \phi=x+\B u(x)$ and $\lambda^3(r,\theta)$. To this end we first take the $v \to 0$ limit in the order $\epsilon$ solution appearing in Eq. (\ref{firstO}) and neglect the T-stress term to obtain \citep{98Fre}
\begin{eqnarray}
\label{QS1st}
u^{(1)}_x(r, \theta)&=&\sqrt{\ell~r}\left[\frac{7}{3}\cos{\left(\frac{\theta}{2}\right)}-
\cos{\left(\frac{3\theta}{2}\right)}\right],\nonumber\\
u^{(1)}_y(r,\theta)&=&\sqrt{\ell~r}\left[\frac{13}{3}\sin{\left(\frac{\theta}{2}\right)}-
\sin{\left(\frac{3\theta}{2}\right)}\right]\ .
\end{eqnarray}
This, together with the second order solution in Eq. (\ref{QS2nd}), yields $\B \phi$ to order $\epsilon^2$. We then use the expansion of $\lambda^3(r,\theta)$ in Eq. (\ref{lambdaExpansion}) to obtain $s_{ij}$ to second order in $\epsilon$. The integrals in Eq. (\ref{spurious}) are then evaluated analytically, yielding
\begin{eqnarray}
\label{spurious1}
f_x &\propto&  \epsilon^2\left(\frac{56}{15}+4a+3b\right) ,\nonumber\\
f_y &=&  0 \ .
\end{eqnarray}
To eliminate unbalanced forces, we demand now that $f_x$ in Eq. (\ref{spurious}) vanishes. This condition, together with Eq. (\ref{a}), yields
\begin{eqnarray}
\label{b}
b=-\frac{52}{45} \ .
\end{eqnarray}
Therefore, the complete weakly nonlinear quasi-static solution for a neo-Hookean material is entirely determined (cf. Eqs. (\ref{solution}), (\ref{particular}), (\ref{QS2nd}) (\ref{a}), (\ref{b}) and (\ref{QS1st})) and reads
\begin{eqnarray}
\label{QS_solution}
&u_x&\!\!\!\!\!\!(r, \theta)=\frac{K_I \sqrt{r}}{4\mu \sqrt{2\pi}}\left[\frac{7}{3}\cos{\left(\frac{\theta}{2}\right)}-
\cos{\left(\frac{3\theta}{2}\right)}\right]\nonumber\\
&+&\!\!\!\!\!\!\left(\frac{K_I}{4\mu \sqrt{2\pi}}\right)^2\left[-\frac{1}{15} \log{\left(r\right)}-\frac{52}{45}\left(\log(r)+\frac{3}{4}\sin^2(\theta) \right) - \frac{103}{48} \cos{(\theta)} +\frac{26}{15} \cos{(2\theta)} - \frac{3}{16} \cos{(3\theta)}\right],\nonumber\\
&u_y&\!\!\!\!\!\!(r,\theta)=\frac{K_I\sqrt{r}}{4\mu \sqrt{2\pi}}\left[\frac{13}{3}\sin{\left(\frac{\theta}{2}\right)}-
\sin{\left(\frac{3\theta}{2}\right)}\right]\nonumber\\
&+&\!\!\!\!\!\!\left(\frac{K_I}{4\mu \sqrt{2\pi}}\right)^2\left[\frac{\theta}{15}-\frac{52}{45}\left(\frac{\theta}{4}-\frac{3}{8}\sin(2\theta) \right)  -\frac{61}{48}\sin{(\theta)} +\frac{26}{15} \sin{(2\theta)} - \frac{3}{16} \sin{(3\theta)}\right] \ ,
\end{eqnarray}
where the only free parameter is $K_I$, in accord with the concept of the autonomy of the near-tip nonlinear zone. The displacement-gradients emerging from this solution still retain a $1/r$ singular contribution. We stress that a solution similar to that of Eq. (\ref{QS_solution}) can be derived for a {\em general} strain energy functional, retaining all of its generic properties. Here we have focused on a neo-Hookean strain energy functional both for concreteness and to enable comparison with the experiments to follow.

There is a key difference between LEFM and the weakly nonlinear theory in relation to the $1/r$ singularity. The displacement fields of Eq. (\ref{QS2nd}) {\em are} a solution of the (LEFM) Lam\'e equation of (\ref{Lame}). The boundary conditions of Eq. (\ref{BC1}) are satisfied by setting $a\!=\!0$. However, in LEFM $f_x$ in Eq. (\ref{spurious}) vanishes if and only if $b\!=\!0$. This implies that the $1/r$ singularity emerging from the gradients of the displacement fields of Eq. (\ref{QS2nd}) must vanish in the framework of LEFM \citep{74Rice}. In contradistinction, the emergence of the inhomogeneous terms proportional to $\B g(\theta;v)$ and $\kappa(v)$ in Eqs. (\ref{secondO})-(\ref{BC2}), within the framework of the weakly nonlinear theory, makes a crucial difference that gives rise to a physically acceptable $1/r$ singular contribution. Thus, not only is this singularity physically acceptable, but it is also {\em necessary}.

Finally, we note that $K_I$ in Eq. (\ref{QS_solution}) cannot be determined by the asymptotic solution, but rather entails either solving the global fracture problem or a direct comparison with an experiment. We will follow the latter in the next section.

\section{Comparison to direct measurements}
\label{comparison}

We now compare the weakly nonlinear solution to the direct measurements of the deformation near a propagating crack tip reported in \citet{08LBF}. Although a similar comparison was performed in \citet{08BLF}, the resulting agreement of the theory with the experimental data was obtained using $B$ of Eq. (\ref{analytic}) as a free fitting parameter. We now test whether the solution appearing in Eq. (\ref{QS_solution}), where there are no free parameters, still agrees with the experimental observations.

\begin{figure}[here]
\centering
\epsfig{width=.45\textwidth,file=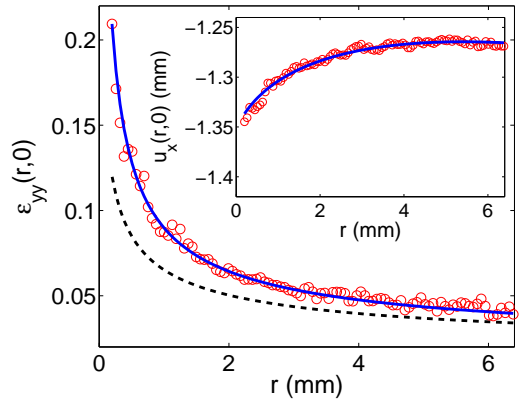}
\caption{(Color online) Inset: The measured $u_x(r,0)$ at $v\!=\!0.2c_s$ (circles) fitted to
the weakly nonlinear solution (solid line) with $K_I\!=\!1040$Pa$\sqrt{m}$ and $T\!=\!-2800$Pa. Main panel: The
corresponding measurements of $\varepsilon_{yy}(r,0)\!=\!\pa_y
u_y(r,0)$ (circles) compared to the weakly nonlinear solution (solid line) using the same $K_I$ and $T$. The LEFM prediction (analysis as in \citet{08LBF}) is added for comparison (dashed line).}\label{fit}
\end{figure}

To perform this comparison, two issues should be addressed. First, the lowest propagation velocity data in the experiments reported in \citet{08LBF} correspond to $v\!=\!0.2c_s$, while the solution in Eq. (\ref{QS_solution}) corresponds to the limit $v \to 0$. We therefore replace the quasi-static order $\epsilon$ part of Eq. (\ref{QS_solution}), i.e. the standard K-field, by its dynamic counterpart for $v\!=\!0.2c_s$ (see Eq. (\ref{firstO})). Second, in these measurements in the same region in space where the order $\epsilon^2$ terms are non-negligible compared to order $\epsilon$ terms, there exists also a non-negligible T-stress contribution \citep{08BLF}. We, therefore, add the T-stress terms appearing in Eq. (\ref{firstO}) (with $\tilde \lambda = 2\mu$ to account for incompressibility) to Eq. (\ref{QS_solution}).

The resulting weakly nonlinear solution, with $K_I\!=\!1040$Pa$\sqrt{m}$ and $T\!=\!-2800$Pa, is compared with the experimental data for $u_x(r,0)$ and $\varepsilon_{yy}(r,0)\!=\!\pa_y
u_y(r,0)$ in Fig. \ref{fit}. The agreement between the weakly nonlinear theory and the experiment is excellent and is a significant improvement to the LEFM result (added to Fig. \ref{fit} for reference). Moreover, the values of $K_I$ and $T$ used here are close (to within 3\% and ~10\%, respectively) to the ones used in Fig. 1(a) of \citet{08BLF}, demonstrating that the experimentally selected free parameter in \citet{08BLF} is consistent with the one determined here theoretically.

\begin{figure}[here]
\centering
\epsfig{width=.45\textwidth,file=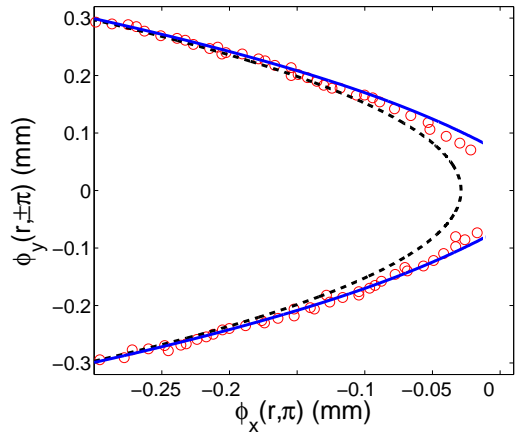}
\caption{(Color online) The measured crack tip profile
($\phi_y(r,\pm\pi)$ vs. $\phi_x(r,\pi)$) at $v\!=\!0.2c_s$ (circles).  Shown are
the parabolic LEFM best fit (dashed line) and the profile
predicted by the weakly nonlinear solution (solid line) with $K_I\!=\!1070$Pa$\sqrt{m}$ and $T\!=\!-2800$Pa.
The values of $K_I$ used here and in Fig. \ref{fit} differ by a few percent, which is consistent with the omission of a correction of order $T/\mu$ mentioned below Eq. (\ref{BC2}).}\label{fit1}
\end{figure}

For completeness, we compare the CTOD predicted by the weakly nonlinear solution with the measured one in Fig. \ref{fit1}. The agreement is favorable and should be compared with Fig. 2(a) in \citet{08BLF}. This result explicitly demonstrates the existence of $\log{r}$ displacement terms that are responsible for the deviation from a parabolic CTOD observed in Fig. \ref{fit1}. We conclude that the weakly nonlinear solution in which only $K_I$ (and $T$ in this particular case) is a free parameter agrees well with the experimental data and validates the previously reported results \citep{08BLF}.

\section{The weakly nonlinear J-integral}
\label{J}

We further substantiate the solution in Eq. (\ref{QS_solution}) by demonstrating the path-independence of the weakly nonlinear J-integral. The dynamic J-integral for nonlinear elasticity reads \citep{80GY, 98Fre}
\begin{equation}
\label{Jint} J=\int_{\C C} \left[\left(U(\B F)+\frac{1}{2}\rho \pa_tu_i
\pa_tu_i\right)v n_x+ s_{ij} n_j\pa_t u_i\right]d{\C C}\ .
\end{equation}
Here ${\C C}$ is a contour that starts at one traction-free face of
the crack, surrounds the tip, ends at the opposite traction-free
face and translates with the crack tip as it moves. $\B n$ is
an outward unit vector on ${\C C}$ and $U(\B F)$ is given in Eq.
({\ref{NH}). This J-integral is path-independent for steady-state
propagation, i.e. when $\pa_t\!=\!-v\pa_x$ \citep{98Fre}. We note that the J-integral of Eq. (\ref{Jint}) is a generalization of the common linear elastic J-integral, where $U(\B F)$ is a general nonlinear strain energy functional, $\B s$ is the first Piola-Kirchhoff stress tensor and $\rho$ is the reference density \citep{98Fre}.
Our aim here is to calculate the integral in Eq. (\ref{Jint}) to {\em third} order in $\epsilon$, which is termed here the weakly nonlinear J-integral.

We are interested in the energy release rate $G=J/v$ under quasi-static conditions, i.e. for $v \to 0$. Therefore we have
\begin{eqnarray}
\label{G}
G \!&=&\! \lim_{v \to 0} \frac{J}{v} \!=\! \int_{-\pi}^{\pi}\!\! \left[\frac{\mu}{2}\left( F_{ij}F_{ij}\!+\!\lambda^{2}\!-\!3\right) n_x \!-\! s_{ij} n_j\pa_x u_i\right]rd\theta\nonumber\\
&\equiv&\epsilon^2 G^{(2)} + \epsilon^3 G^{(3)} + \C O(\epsilon^4) \ ,
\end{eqnarray}
where we choose $\C C$ to be a circle of radius $r \simeq \ell$.
We then expand $\lambda^2$ and $\lambda^3$ to order $\epsilon^3$, expand $s_{ij}$ in Eq. (\ref{sij}) to order $\epsilon^3$ and use Eq. (\ref{QS_solution}) to obtain the integrand of Eq. (\ref{G}) to order $\epsilon^3$. The integration, which involves simple trigonometric functions, can be performed analytically, yielding
\begin{eqnarray}
\label{G1}
G^{(2)} &=&  \frac{32\pi\mu\ell}{3} ,\nonumber\\
G^{(3)} &=&  0 \ .
\end{eqnarray}
Using $\epsilon$ of Eq. (\ref{epsilon}) in Eqs. (\ref{G1})-(\ref{G}) we obtain
\begin{eqnarray}
\label{G2}
G &=&  \frac{K_I^2}{3\mu} \ .
\end{eqnarray}
This is precisely the known LEFM expression \citep{98Fre}.
This result is expected and shows that the weakly nonlinear solution in Eq. (\ref{QS_solution}) is consistent with the path-independence of the J-integral up to order $\epsilon^3$. In fact, we must have $G^{(3)}= 0$ in order for the J-integral to take the same value obtained in the asymptotic LEFM region \citep{98Fre}. Moreover, the order $\epsilon^3$ integrand in Eq. (\ref{G}) (which results in $G^{(3)})$ is more singular than $1/r$, which implies an r-dependence that would have violated the path-independence property if $G^{(3)}\!\ne\! 0$. Thus, again, we must have $G^{(3)}\!=\! 0$, as was obtained explicitly. We conclude that the weakly nonlinear solution appearing in Eq. (\ref{QS_solution}) results in a path-independent weakly nonlinear J-integral, consistent with the path-independence property of the general J-integral of Eq. (\ref{Jint}) under steady-state conditions.

\section{Concluding remarks}
\label{summary}

We demonstrated that the $1/r$ singularity is a necessary and physically acceptable solution in the framework of weakly nonlinear fracture mechanics, in contrast with linear elastic fracture mechanics where such a singularity does not exist. We showed that the resulting weakly nonlinear solution is compatible with the concept of the autonomy of the near-tip nonlinear zone and that the weakly nonlinear J-integral is path-independent, taking the same value as its linear elastic counterpart. These results show that the weakly nonlinear theory is consistent will all of the tenets of fracture mechanics.

In the second part of this work we focused on the quasi-static limit, mainly for the sake of illustration and in order to derive explicit analytical results. The dynamic weakly nonlinear solution is given explicitly in Eqs. (\ref{solution}) and (\ref{solutionA}), where the parameters $A$ and $B$ are related by the boundary conditions of Eq. (\ref{BC2}) \citep{08BLF}. The remaining parameter, say $B$, should be determined by a similar procedure to the one presented in this work. Note, however, that the appearance of a net force transmitted across a line surrounding the crack tip is not necessarily ruled out in the dynamic case ($v\!>\!0$) since this force can be balanced by material inertia (the rate of change of linear momentum). It is clear though, that the condition that the displacement-gradients $1/r$ singularity does not generate any unbalanced force is a necessary and sufficient condition to determine $B$. In \citet{08BLF}, $B$ was determined as a fitting parameter to the experimental data. Here we showed, for the smallest propagation velocity reported in \citet{08LBF, 08BLF}, that the two procedures yielded very similar results, giving us confidence that the results presented in \citet{08BLF} for higher propagation velocities are valid.

The existence of an implicit lengthscale $\ell(v)$ associated with the weakly nonlinear theory may open the way to understand the origin of crack tip instabilities, in line with similar ideas that were proposed recently \citep{03BAG, 06BG}. There is, in fact, some experimental evidence that points in this direction. \citet{07LBDF} reported that, when the micro-branching instability is suppressed in gels, sinusoidal oscillations of a crack's path occur beyond a critical velocity of $0.9c_s$. The wavelength of these high-velocity crack path oscillations is consistent with the mm-scale $\ell(v)$ that arises at the relevant velocity \citep{08BLF}. Moreover, this wavelength was shown to be independent of the sample size \citep{07LBDF}, as opposed to the LEFM-based prediction \citep{07BP}. This observation implies that the wavelength does not originate from the system's geometry, but rather from a dynamical effect. In addition, experimental observations of the micro-branching instability in gels \citep{05LCF} indicated that the typical micro-branch length also seems consistent with $\ell(v)$ in the relevant velocity range.

This set of observations suggests that the weakly nonlinear theory should be seriously considered as a possible route to understand crack tip instabilities. In this regard, we reiterate that the lengthscale $\ell(v)$ marks the breakdown of LEFM and thus the onset of deformation-dependent material behavior. The latter includes, for example, hyperelastic effects such as variable local wave-speeds \citep{96Gao, 03BAG, 06BG} and variable local response times \citep{08BL} that were suggested as a possible instability mechanisms.
The lengthscale $\ell(v)$ is appealing in this context also from the point of view that it is independent of dissipative mechanisms and/or structural lengthscales. This may be important, in view of the existence of identical crack tip
instabilities \citep{99FM, 05LCF} in several amorphous brittle materials with wholly different
dissipative mechanisms and micro-structures.\\

{\bf Acknowledgements}\\

We are grateful to James R. Rice for very fruitful discussions and in particular for introducing us to his 1974 work and for suggesting a way to determine the free parameter in the weakly nonlinear solution. This research was supported by Grant no. 57/07 of the Israel Science Foundation.

\appendix

\section{The expansion of $\lambda^3(r,\theta)$ to second order in $\epsilon$}
\label{appendix}

In this Appendix we present the expansion of $\lambda^3(r,\theta)$, where $\lambda(r,\theta)$ is given in Eq. (\ref{lamb}), to second order in $\epsilon$
\begin{eqnarray}
\label{lambdaExpansion}
\lambda^3(r,\theta)&\simeq& 1-\left[3\left(\cos\theta\pa_r u_x^{(1)}-\frac{\sin\theta}{r}\pa_\theta u_x^{(1)}\right)+3\left(\sin\theta\pa_r u_y^{(1)}+\frac{\cos\theta}{r}\pa_\theta u_y^{(1)}\right) \right]\epsilon\\
&+&\Big[-3\left(\cos\theta\pa_r u_x^{(2)}-\frac{\sin\theta}{r}\pa_\theta u_x^{(2)}\right)-3\left(\sin\theta\pa_r u_y^{(2)}+\frac{\cos\theta}{r}\pa_\theta u_y^{(2)}\right)\nonumber\\
&+&6\left(\cos^2\theta(\pa_r u_x^{(1)})^2+\sin^2\theta(\pa_r u_y^{(1)})^2+\frac{\cos^2\theta}{r^2}(\pa_\theta u_y^{(1)})^2+\frac{\sin^2\theta}{r^2}(\pa_\theta u_x^{(1)})^2 \right)\nonumber\\
&+&6\left(\sin(2\theta)\pa_r u_x^{(1)} \pa_r u_y^{(1)}+\frac{\sin(2\theta)}{r}\pa_r u_y^{(1)} \pa_\theta u_y^{(1)}-\frac{\sin(2\theta)}{r}\pa_r u_x^{(1)} \pa_\theta u_x^{(1)}-\frac{\sin(2\theta)}{r^2}\pa_\theta u_x^{(1)} \pa_\theta u_y^{(1)} \right)\nonumber\\
&+&12\left(\frac{\cos^2\theta}{r}\pa_r u_x^{(1)} \pa_\theta u_y^{(1)} - \frac{\sin^2\theta}{r}\pa_\theta u_x^{(1)} \pa_r u_y^{(1)} \right)+3\left( \frac{1}{r}\pa_\theta u_x^{(1)} \pa_r u_y^{(1)} -\frac{1}{r}\pa_r u_x^{(1)} \pa_\theta u_y^{(1)}\right)\Big]\epsilon^2+\C O(\epsilon^3).\nonumber
\end{eqnarray}

\end{document}